\documentclass[conf]{new-aiaa}
\usepackage[utf8]{inputenc}

\usepackage{graphicx}
\usepackage{amsmath}
\usepackage[version=4]{mhchem}
\usepackage{siunitx}
\usepackage{longtable,tabularx}
\usepackage{amsmath}
\usepackage{color}
\usepackage{soul}
\usepackage{cancel}

\usepackage{algorithm,algorithmicx}
\usepackage{mathtools}
\usepackage[noend]{algpseudocode}
\usepackage{varioref}

\title{Control of laminar boundary-layer separation using steady and harmonic vortex generator jets}

\author{Aria Alimi\footnote{Correspondence: aria.alimi@uni-kassel.de} and Olaf Wünsch}
\affil{Department of Mechanical Engineering, Chair of Fluid Mechanics \\ University of Kassel, 34125 Kassel, Germany}

\begin{document}

\maketitle

\begin{abstract}
\noindent Active flow control of canonical laminar separation bubbles by steady and harmonic vortex generator jets (VGJs) was investigated using direct numerical simulations. Both control strategies were found to be effective in controlling the laminar boundary-layer separation. However, the present results indicate that using the same blowing amplitude, harmonic VGJs were more effective and efficient in reducing the separated region than the steady VGJs considering the fact that the harmonic VGJs use less momentum than the steady case. For steady VGJs, longitudinal structures formed immediately downstream of injection location led to formation of hairpin-type vortices causing an earlier transition to turbulence. Symmetric hairpin vortices were shown to develop downstream of the forcing location for the harmonic VGJs as well. However, the increased control effectiveness for harmonic VGJs flow control strategy is attributed to the fact that shear-layer instability mechanism was exploited. As a result, disturbances introduced by VGJs were strongly amplified leading to development of large-scale coherent structures, which are very effective in increasing the momentum exchange, thus, limiting the separated region.
\end{abstract}

\section{Introduction}
A laminar boundary layer subject to an adverse pressure gradient (APG) is susceptible to separation. In the presence of strong APG, boundary layer will detach from the solid surface. A separated shear-layer exhibits inflectional velocity profiles which support the amplification of small disturbances as a result of fluid dynamics instabilities. The amplified disturbances increase the exchange of momentum and eventually derive the flow to reattach as a turbulent boundary layer, leading to the so-called transitional laminar separation bubbles (LSBs) \cite{1,2}. Laminar separation bubbles can occur in many important aerospace applications operating at low Reynolds numbers, such as low-pressure turbine blades, wings of small unmanned aerial vehicles (UAVs), wind turbine blades and laminar flow airfoils, to name a few. Flow separation can lead to massive degradation of aerodynamic performance characteristics such as loss in lift and a significant increase in drag \cite{3}. Transitional LSBs have been subject of numerous detailed experimental and numerical investigations in that past decades \cite{1,2,3,4,5,6}.

Successful flow control strategies to reduce the negative effects of laminar separation for any aerodynamic vehicles especially operating at low Reynolds numbers could lead to significant performance improvements \cite{7,8,9}. Flow separation control strategies can be classified as either passive or active. The classification is based on whether an external energy is introduced in the flow field or not. Unlike the active flow control (AFC), no direct external energy is required in the passive flow control (PFC) strategy, meaning that the performance of a PFC method is mainly based on mixing high momentum fluid to areas of low momentum to prevent/reduce boundary layer separation \cite{8}. Several approaches have been employed for AFC strategy such as uniform suction and blowing actuation and vortex generator jets (VGJs) through an array of small holes where they are operated either as steady or unsteady/periodic. A full review of the control of flow separation by periodic excitation can be found in \cite{10}. Uniform periodic suction and blowing flow control through a use of slot has been investigated thoroughly and the successful application of this method has been demonstrated \cite{11,12,13,14,15}. The efficiency and effectiveness of periodic/harmonic suction and blowing slot is attributed to exploitation of the fluid dynamics instability associated with the separated shear-layer. In addition to controlling LSBs, uniform blowing has been also shown to be effective in decreasing the skin-friction coefficient for turbulent boundary-layer over a wing section at chord-based Reynolds number of 100,000 \cite{16}.

The present numerical investigation focuses to active separation control using vortex generator jets. VGJs constitute an AFC strategy in which fluid is injected into the boundary layer through an array of small holes. Initially, most of the research regarding VGJs was focused on the application of VGJs on the turbulent boundary layers \cite{17,18}. The use of VGJs for laminar separation control is motivated by the Wind tunnel experiments at the Air Force Research Laboratory (AFRL) at Wright-Patterson Air Force Base \cite{7,9,19,20}. Low pressure turbines (LPTs) are important components of many modern jet engines. The experimental investigations of Bons et al. \cite{19,20} and Sondergaard et al. \cite{7} at AFRL showed that laminar separation from LPT blades for chord Reynolds numbers between 25,000 and 100,000 can be drastically controlled by vortex generator jets. They reported that both steady and pulsed actuation to be effective in reducing separation losses. Whereas pulsed blowing requires a fraction of the momentum compared to the steady VGJs, pulsed forcing was shown to be more efficient than steady case.

One of the first detailed numerical investigations of active flow separation control using pulsed VGJs was direct numerical simulations (DNS) by Postl et al. \cite{21}. Their simulations demonstrated that the effectiveness of pulsed VGJs are based on excitation of an inviscid shear layer instability. It was shown that exploiting the shear layer instability leads to formation of coherent structures which are very effective in reducing the separated region. The effect of free-stream turbulence (FST) on the control of laminar boundary-layer separation using pulsed VGJs was investigated in detail by Hosseinverdi and Fasel \cite{22,23}. Using high fidelity DNS, they found that for low FST levels and a blowing ratio of 0.6, the pulsed jets showed the same effectiveness as observed in zero FST case. When the FST intensity was increased up to 3\%, the effectiveness of the pulsed jets slowly diminished. For FST with 3\% intensity, they investigated the influence of the spanwise jet spacing, the blowing ratio, and the actuation frequency and it was demonstrated that a clear optimum exists for all three parameters. One of their important findings was that the spanwise hole spacing of VGJs plays an important role in the effectiveness of pulsed VGJs flow control for separated flows subjected to high levels of FST intensity.

Active flow control using VGJs can be accomplished by different temporal functions. The main objective of this paper is to investigate and compare the relevant underlying physics of separating control through use of harmonic and steady VGJs actuation, using high resolution DNS. In the next section, Section \ref{num}, the computational methodology including the governing equations and numerical methods are explained. The simulation setup and boundary conditions are presented in Section \ref{setup}. The main results for the uncontrolled LSB and controlled LSBs are discussed in detail in Sections \ref{unc} and \ref{cont}, respectively. A summary and conclusions are provided in Section \ref{conc}.

\section{Numerical Methodology}\label{num}
The following section provides a description of the computational approach used for solving the incompressible Navier-Stokes equations. In this work, the incompressible, three-dimensional unsteady Navier-Stokes equations (N-S) form the set of the governing equations. Instead of solving the N-S equations in primitive variables, the velocity-pressure formulation, the velocity-vorticity formulation of the N-S equations was solved \cite{24} where the transport equation for vorticity vector $\vec{\omega}=(\omega_x,\omega_y,\omega_z)^T$ reads 
\begin{equation}
\frac{\partial\vec{\omega}}{\partial{t}}=(\vec{\omega} \cdot \Delta)\vec{u} - (\vec{u} \cdot \Delta)\vec{\omega}+\frac{1}{Re}\nabla^2 \vec{\omega},
\label{eq:vor}
\end{equation}

In the above equation, the velocities $\vec{u}=(u,v,w)^T$ are non-dimensionalized by the free-stream velocity $U_\infty$, the coordinates are normalized by an arbitrary reference length $L_\infty$, the vorticities $\vec{\omega}$ are non-dimensionalized by $U_\infty/L_\infty$ and the time $t$ is normalized by $L_\infty/U_\infty$. Here, the global Reynolds number is defined as $Re=U_\infty L_\infty/\nu$, where $\nu$ is the kinematic viscosity. 

The wall-normal velocity component, $v$, is obtained by solving the following Poisson equation
\begin{equation}
\frac{\partial^2 v}{\partial^2 x} +\frac{\partial^2 v}{\partial y^2} + \frac{\partial^2 v}{\partial z^2}=\frac{\partial\omega_z}{\partial x}-\frac{\partial\omega_x}{\partial z}.
\label{eq:v}
\end{equation}
The remaining velocity components, $u$ and $w$, are calculated using the definition of the spanwise and streamwise vorticity components,
\begin{equation}
\begin{aligned}
\frac{\partial u}{\partial y}&=\frac{\partial v}{\partial x} - \omega_z, \\
\frac{\partial w}{\partial y}&=\frac{\partial v}{\partial z} + \omega_x.
\label{eq:u_w}
\end{aligned}
\end{equation}

The governing equations are solved in a three-dimensional rectangular domain using high-order finite difference methods. All spatial derivatives including convective and viscous terms in the streamwise and wall-normal directions are approximated with standard fourth-order compact difference schemes \cite{25}, except for the first derivative of non-linear terms in the $x$-direction. It is known that application of central compact difference discretizations to high Reynolds number flows typically leads to numerical instability. It is attributed to the accumulation of the aliasing errors resulting from discrete evaluation of the nonlinear convective terms in the $x$-direction, owing to the fact that central compact difference schemes are characterized by zero numerical dissipation error and therefore, are prone to aliasing error. A common practice to remove aliasing errors and hence enhancing the numerical stability is filtering the numerical solution at every time level. Instead, an upwind combined compact difference (CCD) scheme proposed in \cite{26} was employed. They demonstrated that their proposed upwind CCD scheme has non-zero dissipation error (numerical diffusion) restricted to the high wavenumber region, while exhibiting very good spectral resolution characteristic. The flow field is assumed to be periodic in the $z$-direction allowing the flow field to be expanded in Fourier cosine and sine series. A fourth-order accurate Runge–Kutta scheme is used to integrate the vorticity transport equations in time.

The computational time needed to solve the N-S in the velocity-vorticity formulation is dominated by the numerical solution of the $v$-velocity Poisson equation, Equation (\ref{eq:v}). In this work, a fourth-order discretization method and an efficient solution algorithm, proposed in \cite{26,27} to solve the Poisson equation, was employed for solving Equation (\ref{eq:v}). Furthermore, a fourth-order accurate compact scheme was employed to obtain $u$ and $w$ velocity according to Equation (\ref{eq:u_w}). At each stage of the Runge–Kutta integration, the vorticity transport equations, Equation (\ref{eq:vor}), are used to advance the vorticity vector over one Runge-Kutta step. Next, the Poisson equation for the wall-normal velocity, Equation (\ref{eq:v}), is solved at the new time level. Then, the streamwise and spanwise velocity components, Equation (\ref{eq:u_w}), are computed at the new time level.
 
\section{Computational Setup}\label{setup}
\begin{figure}[htbp]
    \centering
      \begin{minipage}{4.75in}(\textbf{a})\end{minipage}\\
      \includegraphics[width=0.7\textwidth]{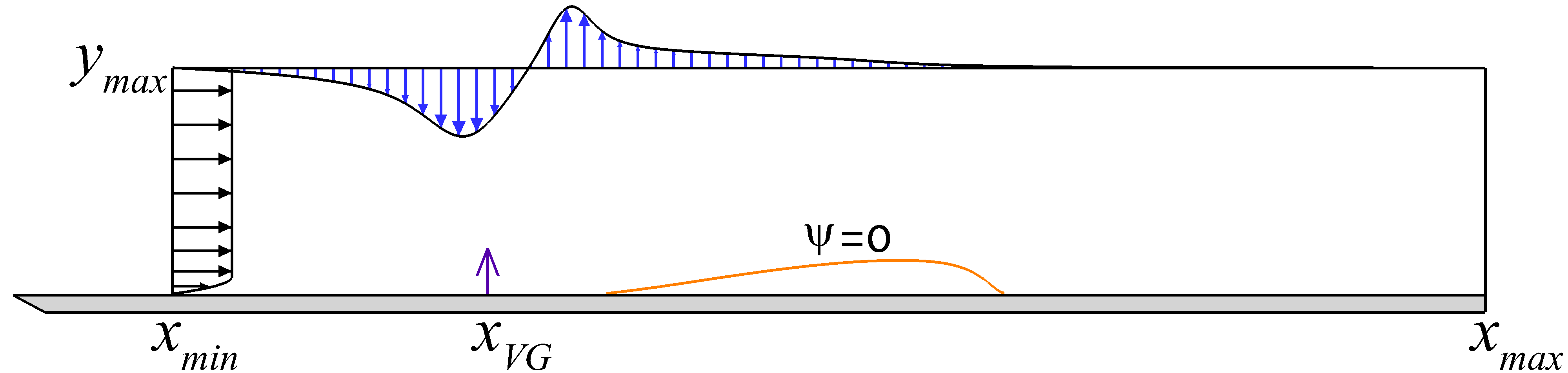}\\
      \begin{minipage}{4.75in}(\textbf{b})\end{minipage}\\
      \includegraphics[width=0.7\textwidth]{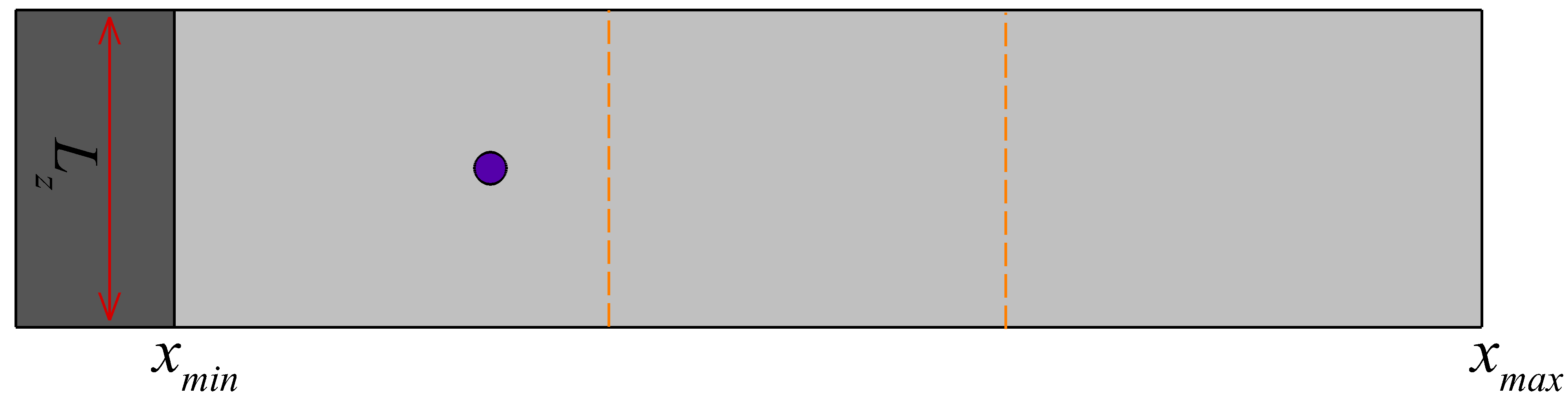}
    \caption{Schematic of the computation setup in side (\textbf{a}) and top-down (\textbf{b}) views (drawing not to scale). The integration domain does not include the flat-plate leading edge. Included in (\textbf{a}) are the wall-normal velocity distribution applied at the free-stream boundary at $y=y_{max}$ and the velocity profile at the inflow boundary. In (\textbf{a}), $x_{VG}$ indicates the streamwise location of vortex generator jets at the wall (up arrow) and the hole in (\textbf{b}) indicates the spanwise location of VGJs. Also, shown in (\textbf{a}) is the typical mean dividing streamline for an uncontrolled LSB and its corresponding separation and reattachment locations as shown in (\textbf{b}).}
    \label{fig:setup}
\end{figure}

The key features of simulation set-up and boundary conditions are explained in this section. The integration domain is a three-dimensional rectangular section. A sketch of this domain can be seen in Figure \ref{fig:setup}. The streamwise coordinate is $x$, the direction normal to the flat-plate is $y$ and the spanwise coordinate is $z$. In the simulations, a wall-normal velocity distribution at the free-stream boundary of the computational domain was specified to create a favorable-to-adverse pressure gradient. The adverse pressure gradient induces a laminar separation bubble on the flat-plate. This simulation setup is defined such that it exhibits a laminar boundary layer separation with similar characteristics as the flow separation over an airfoil (at low angle of attack) but at reduced geometric complexity. The flat-plate model geometry has been extensively employed for investigation of transition in LSBs and separation control in canonical separation bubbles \cite{2,5,6,11,15,28,29,30,31}.

Domain extent and grid resolution used in the present work was guided by the direct numerical simulations of LSBs in \cite{14,15,30}. In the streamwise direction, 1601 grid points are equidistantly distributed in the range $x_{min}=5\leq x\leq x_{max}=19.4$. An exponential grid point distribution was employed in the $y$ direction with 240 grid points. The free-stream boundary of the computational domain is located at $y_{max}=2$.  The homogenous spanwise direction ($z$-direction) was resolved with 127 Fourier modes (200 collocation points) for the domain width of $L_z=2$. The coordinates and velocities were made dimensionless with the same length and velocity scales as used in \cite{15,30}, i.e. $L_\infty=0.0254[m]$ and $U_\infty=6.64[m/s]$, respectively.

The streamwise and spanwise grid spacings in wall units are $\Delta x^+=5.5$ and $\Delta z^+=6.1$, respectively. The wall-normal grid resolution in wall units adjacent to the wall is $\Delta y_w^+=0.9$. The grid resolutions in wall units were computed based on the maximum skin friction coefficient in the turbulent boundary layer downstream of the reattachment point. The grid resolution in the present simulations is comparable or finer compared to the resolutions used in the available literature \cite{4,28,29,30}.

A Blasius solution with the displacement thickness Reynolds number of $Re_{\delta^*}=407$ is prescribed as Dirichlet conditions at the inflow boundary. At the free-stream boundary, a wall-normal velocity distribution is applied at the upper boundary for generating the streamwise pressure gradients which is identical to the those used by \cite{14,15,30}. It should be noted that $v$-velocity at the upper boundary is uniform in the $z$-direction. Furthermore, the wall normal derivatives of vorticity components are set to zero. The no-slip and no-penetration conditions are enforced on the surface of the flat-plate, except at a location where disturbances associated with the flow control are introduced. At the spanwise boundaries of the domain, periodic boundary conditions are applied for all variables and their derivatives. All second derivatives in the streamwise direction are set to zero at the outflow boundary. Furthermore, convective boundary conditions are applied at the outflow boundary \cite{31}.

\section{Uncontrolled flow: Main Features and Validation}\label{unc}
In this section, results obtained from 3-D DNS for the uncontrolled (natural, i.e. no external disturbances) LSB are presented in order to understand the main characteristics of the uncontrolled case. Furthermore, the results obtained from DNS of the natural LSB are compared with the available DNS data \cite{14,15,30} for validation of numerical methods and computational setup.

\begin{figure}[htbp]
    \centering
      \begin{minipage}{5.in}(\textbf{a})\end{minipage}\\
      \includegraphics[width=0.75\textwidth]{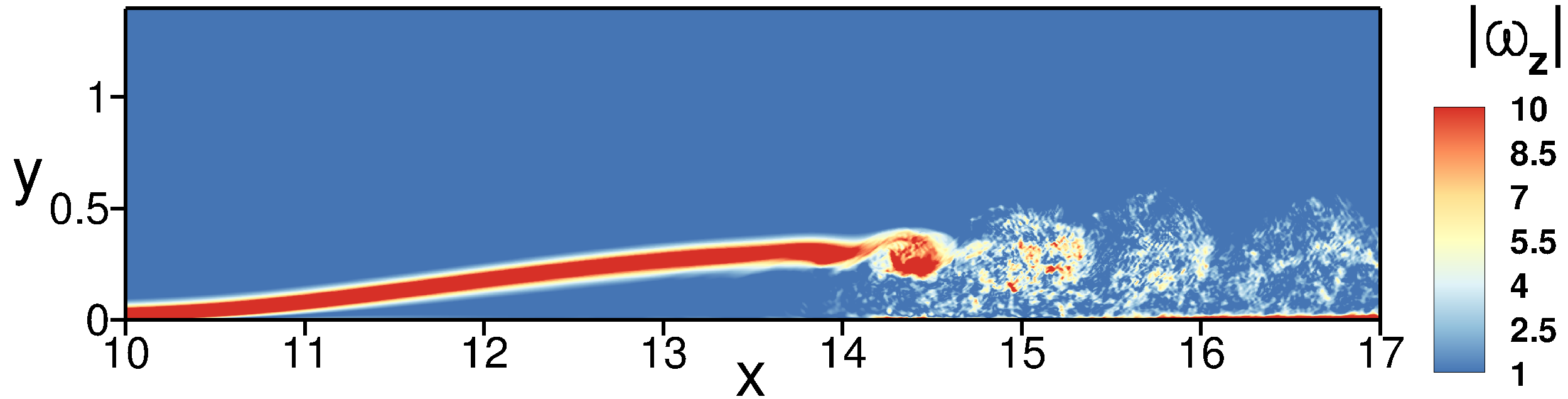}\\
      \begin{minipage}{5.in}(\textbf{b})\end{minipage}\\
      \includegraphics[width=0.75\textwidth]{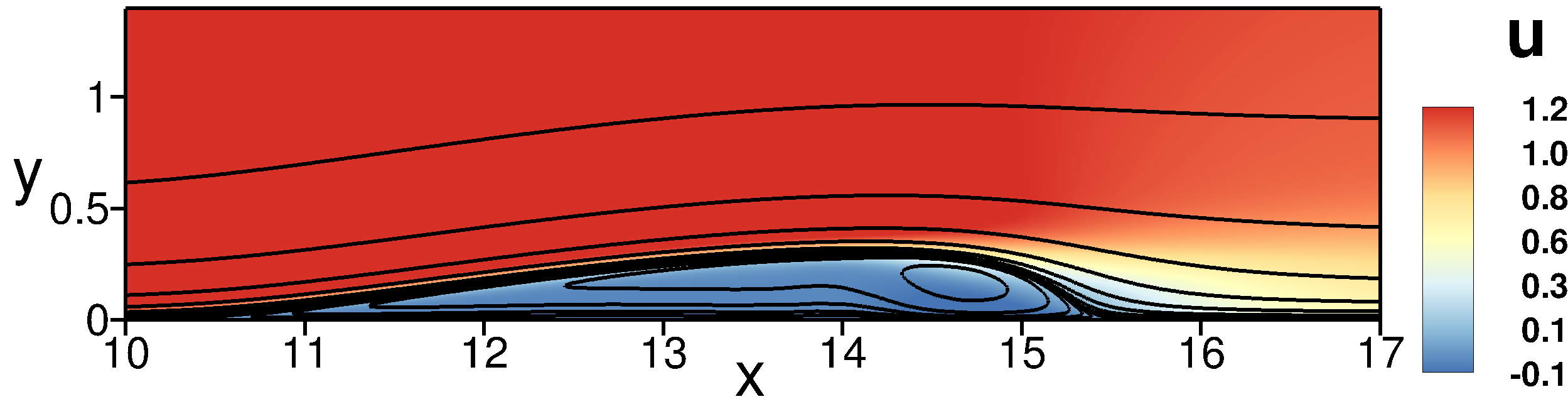}
    \caption{(\textbf{a}) Instantaneous spanwise-averaged spanwise vorticity. (\textbf{b}) Time and spanwise averaged streamline colored with $u$-velocity.}
    \label{fig:unc_inst}
\end{figure}

The instantaneous contours of spanwise-averaged $\omega_z$-vorticity (a random snapshot after the flow reaches the statistically stationary state) is presented in Figure \ref{fig:unc_inst}a to gain insight into the nature of the unsteady flow field of the uncontrolled LSB. To have a complete picture about the mean features of the LSB, the time- and spanwise-averaged streamline together with the contours of mean $u$-velocity are shown in Figure \ref{fig:unc_inst}b. The laminar boundary layer separates at $x_s=10.4$ followed by a smooth separated shear layer with negligible dependency in the spanwise direction. Downstream of the separation location, the separated laminar shear layer contains inflectional reverse-flow velocity profiles where the normal gradient of the $u$-velocity is maximum near the inflection points, therefore, the $\omega_z$-vorticity attains its maximum value at the same wall-normal location at a given $x$-location. Initially, the inflection point remains close to the surface, so the viscous instability plays a more important role. Moving further downstream, the thickness of the separated shear layer increases and the streamwise velocity profiles become very similar to those of free shear layer. As a result of the large growth rates associated with the separated shear layer instability, the disturbances (originated from numerical and round-off errors) can be strongly amplified and lead to the development of large unsteady vortical structures.

The instantaneous flow visualization in Figure \ref{fig:unc_inst}a indicates that that farther downstream the shear layer is wavy and rolls up into spanwise vortices, which detach from the shear layer and are convected downstream, the so-called vortex shedding process. Observed periodic shedding of spanwise coherent vortical structures are the indication of the fact that fluctuating disturbances reach large (non-linear) amplitudes within the separated region \cite{6}. The presence of the large-amplitude disturbance waves which is manifested by large vortices, considerably increases the wall-normal momentum exchange by transporting low-momentum fluid away from the wall and high-momentum free-stream fluid towards the wall. Thus, this increased momentum exchange limits the extent of the separation, which in the time-average picture as shown in Figure \ref{fig:unc_inst}b, it is indicated by the appearance of a reverse-flow vortex and the reattachment.

\begin{figure}[ht]
\centerline{
\noindent
\begin{minipage}{.5\textwidth}
\quad\quad\quad\quad\quad\quad\quad\quad\quad\quad\quad\quad(\textbf{a})
\end{minipage}
\begin{minipage}{.5\textwidth}
\quad\quad\quad\quad\quad\quad\quad\quad\quad\quad\quad\quad\quad(\textbf{b})
\end{minipage}}
\centerline{
      \includegraphics[width=0.52\textwidth]{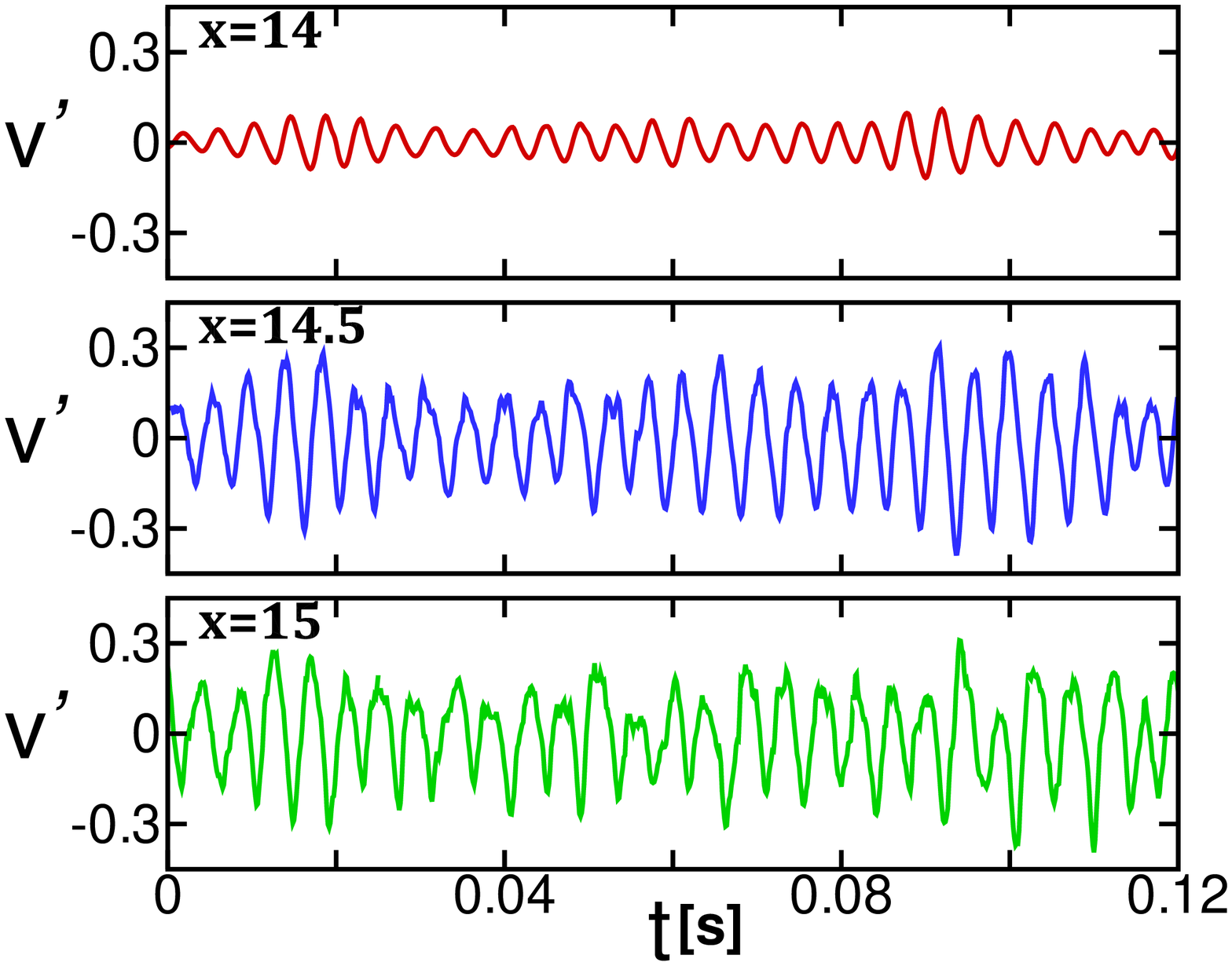}
      \includegraphics[width=0.48\textwidth]{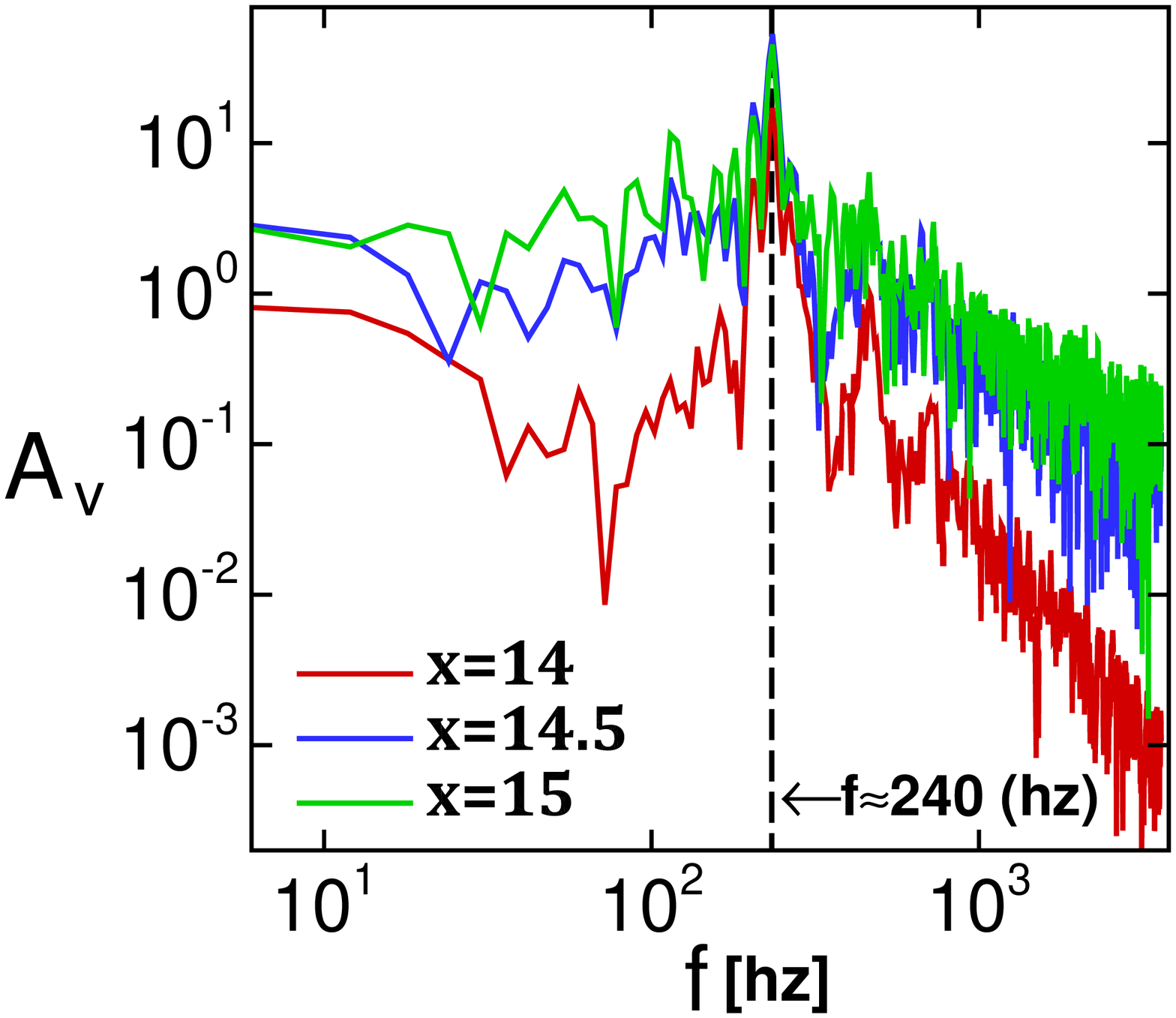}}
    \caption{Temporal evolution  (\textbf{a}) and the corresponding frequency spectra  (\textbf{b}) of the wall-normal disturbance velocity (averaged in spanwise direction) at various streamwise locations along the centerline of separated shear layer.}
    \label{fig:unc_spec}
\end{figure}
\begin{figure}[hbtp]
\centerline{
\noindent
\begin{minipage}{.5\textwidth}
\quad\quad\quad\quad\quad\quad\quad\quad\quad\quad(\textbf{a})
\end{minipage}
\begin{minipage}{.5\textwidth}
\quad\quad\quad\quad\quad\quad\quad\quad\quad(\textbf{b})
\end{minipage}}
\centerline{
      \includegraphics[width=\textwidth]{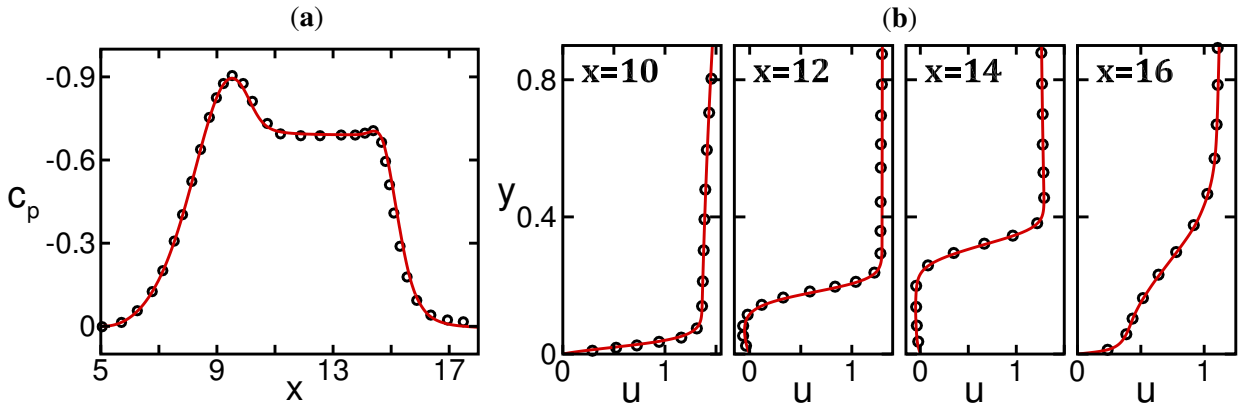}}
    \caption{Time- and spanwise-averaged wall-pressure coefficient (\textbf{a}) and wall-normal profiles of u-velocity (\textbf{b}). Line: present numerical results; symbols: DNS data reported by \cite{30}.}
    \label{fig:unc_comp}
\end{figure}

It is worth noting that the Reynolds number based on the free-stream velocity and the mean separation region ($l$), the streamwise distance between mean reattachment and separation point, is $Re_l=57,793$. It is important to note that the boundary layer developing downstream of the reattachment location is not an equilibrium turbulent boundary layer which is consistent with the finding in \cite{32} as they showed that the boundary layer subjected to a non-constant pressure gradient appears to converge towards the canonical state after a sufficiently long downstream length.

To have a better understating of the unsteady nature of the LSB and estimate the dominant frequency associated with the vortex shedding, the time traces and the respective frequency spectra of the wall–normal disturbance velocity (averaged in spanwise direction) are provided in Figure \ref{fig:unc_spec} for several streamwise locations within the bubble. The data are extracted along the local displacement thickness which is very close to the center of shedding vortex core. All spectra show the development of a distinct peak close to $f = 240~(hz)$, which corresponds to the shedding frequency of the uncontrolled separation bubble. This finding is consistent with the results of the LSB reported in \cite{14,15,30}. 

Prior to the detailed discussion of the investigations for the VGJ controlled LSBs, the mean flow results obtained from the present numerical simulations for the uncontrolled case are compared to the results from DNS reported by \cite{30} in Figure \ref{fig:unc_comp}. Shown are the streamwise distributions of the wall-pressure coefficient, $c_p$, and wall-normal profiles of the streamwise velocity. In Figure \ref{fig:unc_comp}, the velocity profiles were selected from upstream of the separation location, inside the separated region and downstream of the mean reattachment location for comparison purposes. The present DNS results show excellent agreement with those reported by \cite{30}, which confirms the proper implementation and accuracy of the numerical methods and the computational setup in this work.

\section{Controlled Flow: Steady and Harmonic VGJs}\label{cont}
This section is devoted towards understanding the response of laminar boundary layer separation to various vortex generator jets flow control strategies. Before proceeding further, a numerical procedure to model VGJs is explained. Vortex generator jets are realized by specifying appropriate boundary condition for the velocity at the wall. In particular, a non-zero wall boundary condition for the wall-normal velocity is prescribed to simulate VGJs as
\begin{equation}
    v_{VGJ}(x,z,t)=A_{max}~cos^3\bigg(\pi \frac{|\vec{r}-\vec{r}_{VG}|}{D}\bigg)~G(t),
    \label{eq:v_w}
\end{equation}
where $A_{max}$ is the maximum forcing amplitude (blowing amplitude) and $G(t)$ is a function to provide the temporal behavior of the controlled disturbance input. This approach was implemented successfully in DNS of pulsed vortex jets flow control \cite{22,23}. In Equation (\ref{eq:v_w}), $D$ is the hole diameter. It is important to note that the forcing function is only applied on grid points at wall for which $|\vec{r}-\vec{r}_{VG}|<D/2$ where $r=(x,0,z)^T$ and $\vec{r}_{VG}=(x_{VG},0,0)^T$ defines the center of the hole. For all cases, VGJ hole had a diameter of $D=0.2$ and were located at $x_{VG}=9$, which is near the onset of APG. It is worth noting that because of the enforced spanwise periodicity of the simulations, a spanwise VGJ spacing is identical to the spanwise domain width. As explained in \cite{22,23}, the main reason why $cos^3$ velocity distribution was chosen, is the fact that Equation (\ref{eq:v_w}) provides a smooth Poiseuille-type velocity profile for the jet exit velocity and thus avoids large spatial gradients near the edge of hole.

\begin{figure}[hbtp]
\centerline{
\noindent
\begin{minipage}{.5\textwidth}
\quad\quad\quad\quad\quad\quad\quad\quad\quad\quad\quad\quad\quad(\textbf{a})
\end{minipage}
\begin{minipage}{.5\textwidth}
\quad\quad\quad\quad\quad\quad\quad\quad\quad\quad\quad(\textbf{b})
\end{minipage}}
\centerline{
      \includegraphics[width=\textwidth]{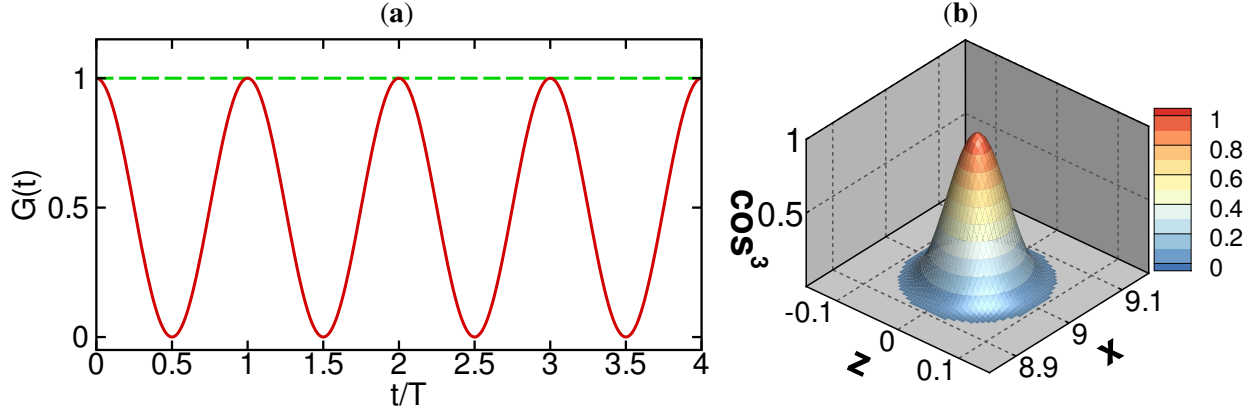}}
    \caption{(\textbf{a}) ) Reproduced time-signal of two different forcing strategy: harmonic (solid red line) and steady (dashed green lines) VGJs. (\textbf{b}) Spatial distribution of the jet velocity at the wall. In (\textbf{a}), $T$ represents the fundamental forcing period, $T=1/F$.}
    \label{fig:v_w}
\end{figure}

In this investigation, two different forcing strategies were considered: (i) a harmonic blowing $G(t) =0.5[1+sin(2\pi Ft)]$ and (ii) steady injection $G(t)=1$. For the harmonic forcing, $F$ is the fundamental forcing frequency and it is set to $F=240~(hz)$ which is the dominant natural shedding frequency of the unforced separation bubble. Figure \ref{fig:v_w} presents the time signal of the wall-normal velocity at the forcing location for different forcing strategies as well as the spatial distribution of the jet velocity at the wall.

All simulations were carried out using the same maximum amplitude of $A_{max}=0.3$. A momentum coefficient, $c_\mu$, of VGJs is proportional to the integral of the square of the time signal over one forcing period, $c_\mu\propto\int{G^2(t)dt}$. Therefore, different forcing strategies lead to different momentum coefficients. The corresponding momentum coefficients for the steady and harmonic cases are $c_\mu\approx8.37\times10^{-5}$ and $c_\mu\approx6.28\times10^{-5}$, respectively, which are in the same range employed in pulsed VGJs investigations \cite{21,22,23}.

\begin{figure}[hbtp]
    \centering
    \includegraphics[width=\textwidth]{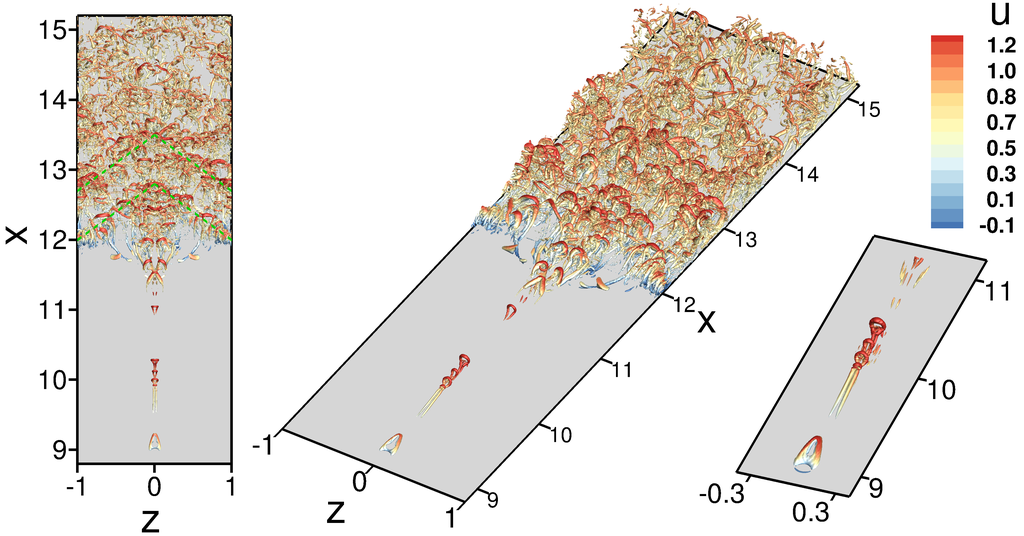}
    \caption{Instantaneous flow visualizations for controlled LSB using harmonic VGJs. Plotted are iso-surfaces of $\lambda_2=-50$ colored by $u$-velocity in top-down and perspective views. Close-up view (right plot) represents the flow structure near the actuator based on $\lambda_2=-20$. Dashed lines in the left plot indicate the generation of oblique coherent structures.}
    \label{fig:inst_hvgj}
\end{figure}

\begin{figure}[htbp]
    \centering
    \includegraphics[width=\textwidth]{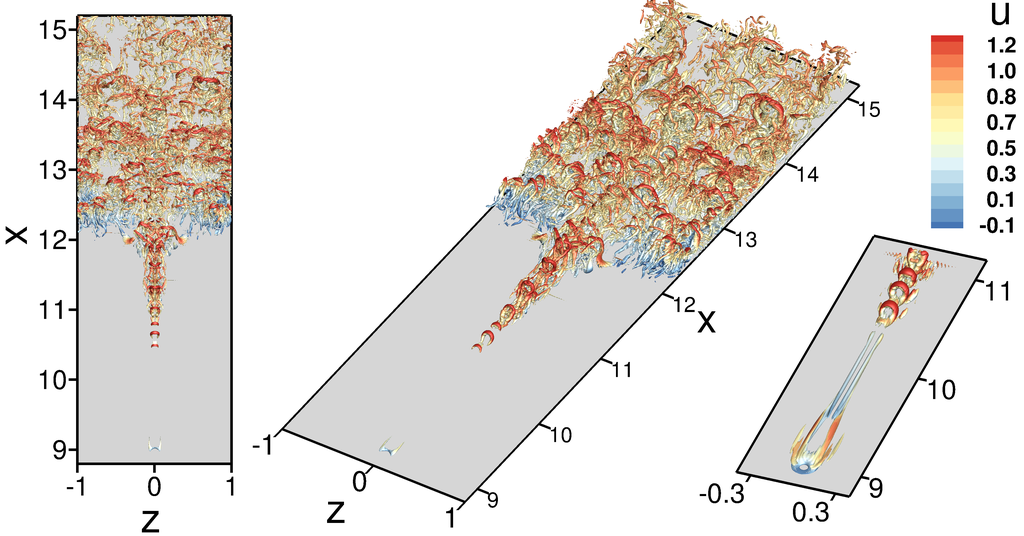}
    \caption{Instantaneous flow visualizations for controlled LSB using steady VGJs. Plotted are iso-surfaces of $\lambda_2=-50$ colored by $u$-velocity in top-down and perspective views. Close-up view (right plot) represents the flow structure near the actuator based on $\lambda_2=-20$.}
    \label{fig:inst_svgj}
\end{figure}

The fundamental physical mechanisms associated with separation control using harmonic and steady VGJs are investigated next. The $\lambda_2$ vortex identification \cite{33} provides insight into the underlying flow physics of controlled LSBs and nature of the vortical structures and their evolution. The $\lambda_2$-criterion is based on a decomposition of the velocity gradient tensor, $\partial u_i/\partial x_j$,  into the symmetric part (the rate-of-strain-rate tensor, $S_{ij}$) and the antisymmetric part (the vorticity tensor, $\Omega_{ij}$). Consider the three real eigenvalues ($\lambda_1\leq\lambda_2\leq\lambda_3$) of the symmetric tensor $S_{ik} S_{kj}+\Omega_{ik} \Omega_{kj}$, a vortex region is defined where $\lambda_2<0$ according to Jeong and Hussain \cite{33}. Figures \ref{fig:inst_hvgj} and \ref{fig:inst_svgj} present the instantaneous iso-surfaces of the $\lambda_2$-criterion colored by streamwise velocity in top-down and perspective views for harmonic and steady VGJs, respectively. A close-up perspective view of the flow structures immediately downstream of the actuator is presented in Figures \ref{fig:inst_hvgj} and \ref{fig:inst_svgj} (right plot) as well. These structures are associated with the high-amplitude forcing through the jet hole.

For harmonic VGJs shown in Figure \ref{fig:inst_hvgj}, downstream of the hole a pair of counter-rotating longitudinal vortices is seen to develop close to the wall. Traveling further downstream, these vortices stretch in the wall-normal direction leading to the formation of hairpin-type vortices. The hairpin vortices lift up from the surface and appear to trigger transition to turbulence which manifested by the emergence of $\Lambda$-structures around $x=11.5$ and $z=0$. An important observation is the formation of oblique large-scale coherent structures in the separated shear layer, which are marked by green dashed lines in the left plot in Figure \ref{fig:inst_hvgj}. The generation of these organized structures could be an indication of excited instability waves, which will be discussed later. 

For steady VGJs case presented in Figure \ref{fig:inst_svgj}, longitudinal structures are developed immediately downstream of the jet hole, where the symmetric counter-rotating vortex pair remains intact for some streamwise distances. Further downstream, the iso-surfaces clearly illustrate the formation of hairpin like vortices which initiate the breakdown process to turbulence. One of the main differences between the steady VGJs forcing with the harmonic one is the lack of organized 2-D or 3-D coherent structures. It appears that the flow reattachment is caused by the increased momentum exchange due to the longitudinal vortices and accelerated transition to turbulence.


\begin{figure}[htbp]
    \centering
    \includegraphics[width=0.6\textwidth]{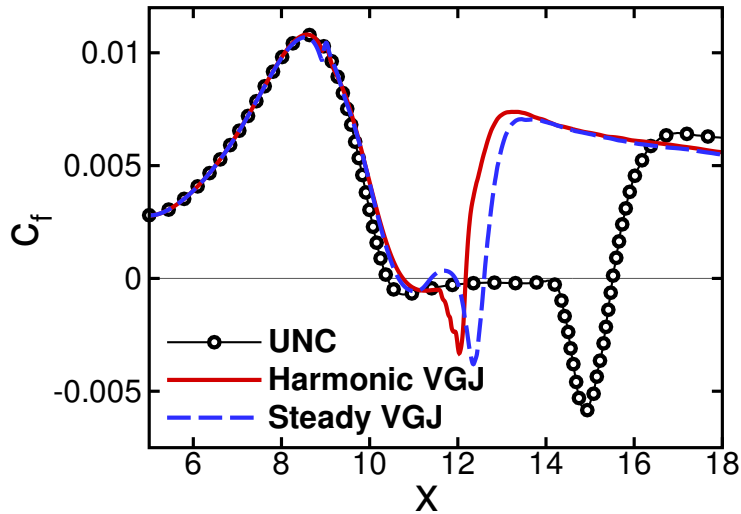}
    \caption{Comparison of time- and spanwise-averaged wall skin friction coefficient for uncontrolled and controlled LSBs obtained from different forcing strategies. UNC refers to the uncontrolled case.}
    \label{fig:cf}
\end{figure}

\begin{figure}[htbp]
    \centering
    \includegraphics[width=0.55\textwidth]{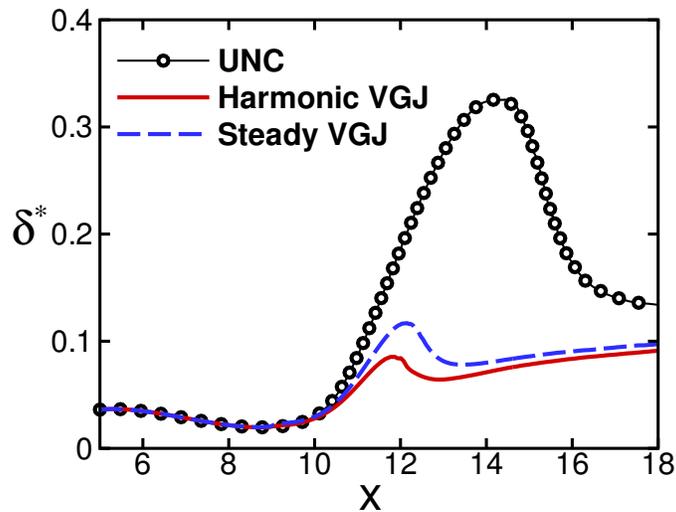}
    \caption{Comparison of time- and spanwise-averaged displacement thickness for uncontrolled and controlled LSBs obtained from different forcing strategies. UNC refers to the uncontrolled case.}
    \label{fig:disp}
\end{figure}

The effectiveness of the vortex generator jet actuation for all cases can be evaluated by comparing time- and spanwise-averaged results. Figures \ref{fig:cf} and \ref{fig:disp} present a comparison of the skin-friction coefficient, $c_f$, and displacement thickness, $\delta^*$, for the uncontrolled and controlled cases, respectively. While the separation length is significantly reduced using different VGJs, the steady VGJs is less effective in reducing the separated region compare to the harmonic case as shown in Figure \ref{fig:cf}. In particular, the mean separation length is reduced by 72\% and 62\% for the harmonic and steady VGJs, respectively. Here, the mean separation length is defined as the distance between the reattachment point at which $c_f$ changes from negative to positive and the separation location, where $c_f$ changes from positive to negative. It is worth noting that the separation location is delayed for both controlled cases with the furthest separation location for the harmonic case. The displacement effect of the separation bubble manifests itself in strong increases in the boundary-layer displacement thickness, where $\delta^*$ is defined as
\begin{equation}
\delta^*=\int_0^{\delta}\big(1-\frac{u}{U_{psd}}\big)dy,
    \label{eq:disp}
\end{equation}
where $\delta$ is the local boundary-layer thickness and  $U_{psd}$ is a `pseudo' free-stream velocity, which is obtained by a wall-normal integration of the spanwise vorticity \cite{29}. From the displacement thickness distribution in Figure \ref{fig:disp}, it is found that in addition to reduction in separation length, all forcing strategies were found to be very effective in reducing the wall-normal extent of the separated region. Here, the steady actuation is less effective in reducing the bubble height.

The boundary layer separation in uncontrolled case is almost two-dimensional and the separated region can be universally identified by the vanishing of wall shear. However, flow separation in controlled LSBs using VGJs can be distinctly different from two-dimensional separation due to the localized (3-D) effects of flow structures associated with VGJs. Therefore, for a better understanding of the mean flow topology near the wall, it is instructive to illustrate the skin-friction lines for the different controlled cases as shown
\begin{figure}[ht]
    \centering
    \includegraphics[width=0.6\textwidth]{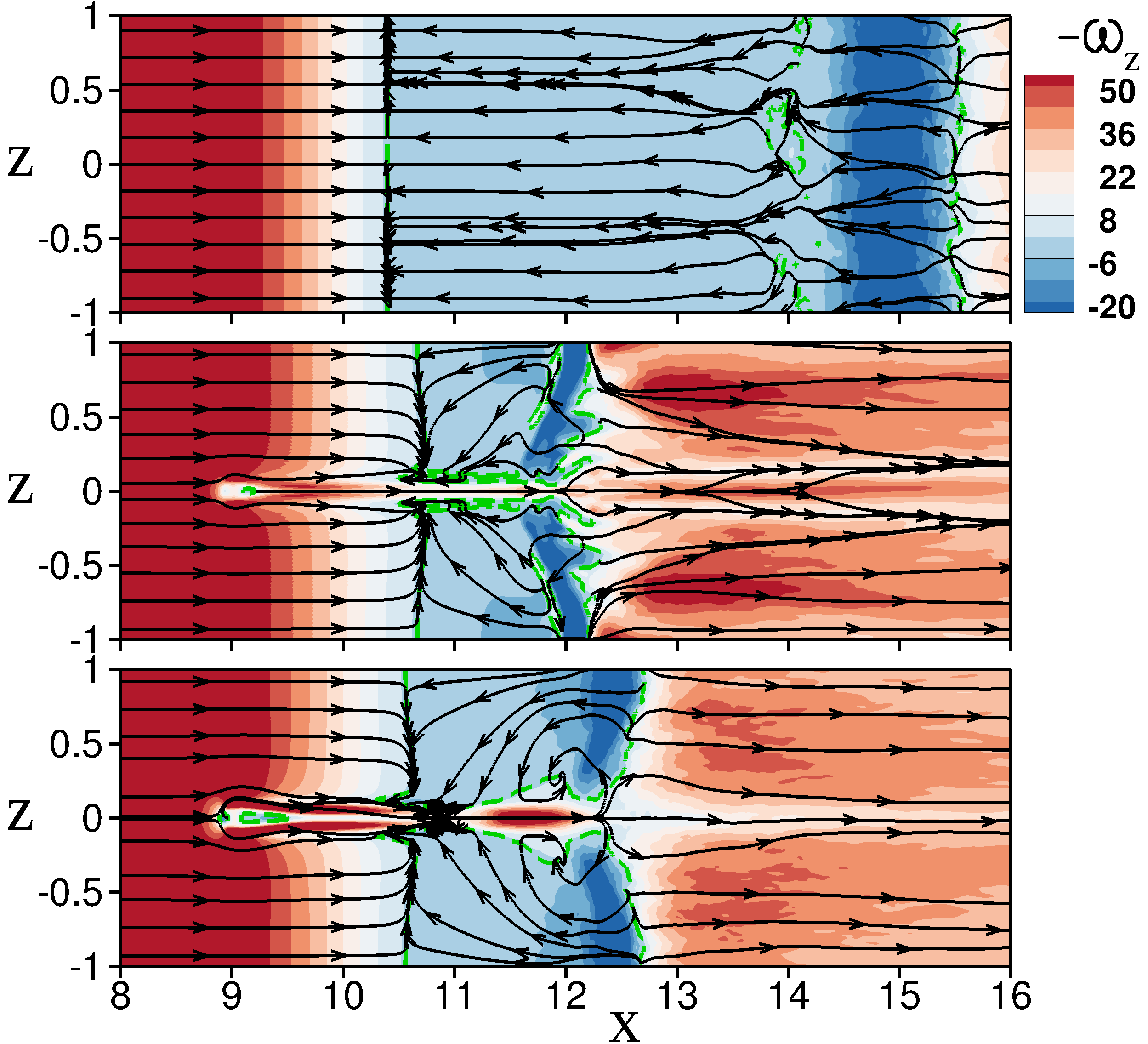}
    \caption{Skin friction lines and color contours of spanwise vorticity on the flat-plate obtained for uncontrolled and controlled LSBs. From top to bottom: uncontrolled, harmonic and steady VGJs. Dashed lines correspond to the time-averaged $\omega_z=0$.}
    \label{fig:sfl}
\end{figure}
in Figure \ref{fig:sfl}. The skin-friction lines are constructed using the streamwise and spanwise wall-vorticity components. In Figure \ref{fig:sfl}, the skin friction lines are colored by the wall-spanwise vorticity together with time averaged $\omega_z=0$ (dashed lines). Note that $\omega_z$-vorticity at the wall are directly proportional to the local skin friction. The results from the uncontrolled case is also included in Figure \ref{fig:sfl} for comparison. In the uncontrolled case, the skin-friction lines converging in a single line are an indication of a singularity in the flow field, that is separation line. On the other hand, skin friction lines diverging from a single line show the location of flow reattachment. The effect of the VGJs on the mean flow pattern near the wall is immediately visible around $x=9$, which corresponds to the forcing location. The localized effect is related to the fact that vortex generator jets were employed in the wall-normal direction. For the controlled LSBs, the separation line is no longer straight in the $z$-direction and strongly modulated in the spanwise direction. Another interesting observation is that the boundary layer remains attached in the plane $z=0$ even downstream of the separation line. Whereas skin-friction lines are nearly parallel downstream of the separation line for the natural LSB, they exhibit complex three-dimensional pattern for controlled cases.

\begin{figure}[ht]
\centerline{
\noindent
\begin{minipage}{.5\textwidth}
\quad\quad\quad\quad\quad\quad\quad\quad\quad\quad\quad\quad\quad\quad(\textbf{a})
\end{minipage}
\begin{minipage}{.5\textwidth}
\quad\quad\quad\quad\quad\quad\quad\quad\quad(\textbf{b})
\end{minipage}}
    \centerline{
    \includegraphics[width=0.9\textwidth]{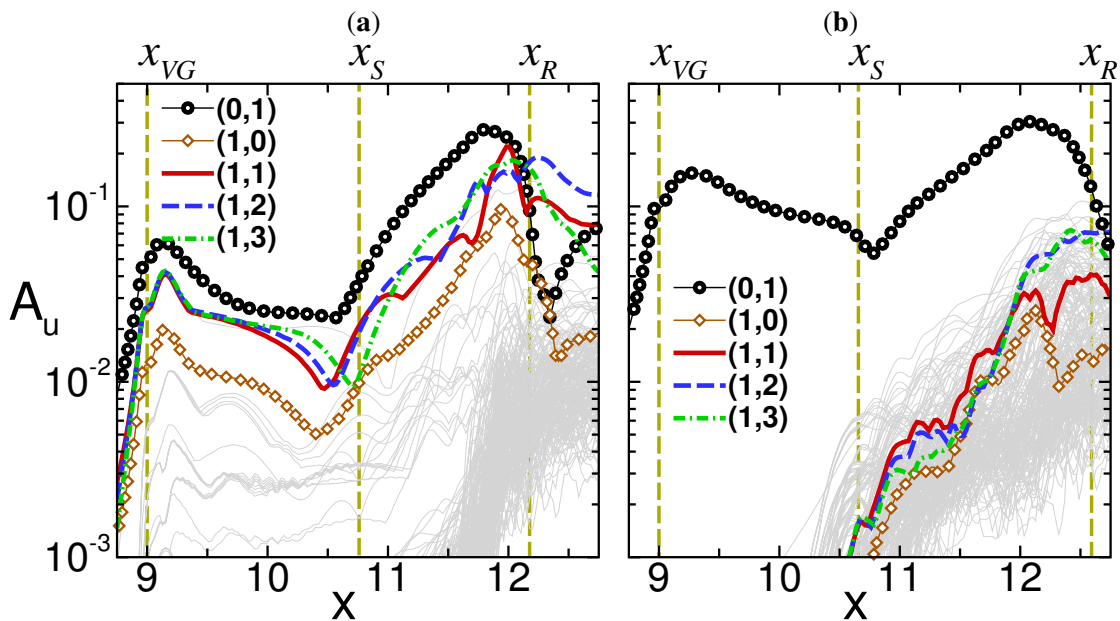}}
    \caption{Disturbance development of the Fourier modes of controlled flow with different forcing strategies in ($n,m$) notation. Plotted are the maximum Fourier amplitude of disturbance $u$-velocity. (\textbf{a}) Harmonic, (\textbf{b}) steady VGJs. The vertical dashed lines correspond to the forcing location, mean separation and reattachment locations $(x_{VG},x_{s},x_{R})$.}
    \label{fig:Au}
\end{figure}

The instantaneous flow structures visualizations, as shown in Figures \ref{fig:inst_hvgj} and \ref{fig:inst_svgj}, do not provide a complete understanding of the underlying flow physics of different VGJs flow control strategies. The question about why steady VGJs is less effective than harmonic actuation in controlling the separation bubbles, despite the fact that it injects more momentum into the flow field, remains unanswered. To identify the dominant mechanisms for different flow control strategies investigated, the time-dependent velocity field were Fourier decomposed in time and spanwise direction. The double Fourier decomposition was motivated by DNS of pulsed VGJs in \cite{23}, where they successfully employed this approach to understand and identify the underlying flow physics of pulsed VGJs flow control. For Fourier transform in time, the flow data are sampled with a time interval of $\Delta t=T/40$ for 10 forcing periods where $T$ is the period of the actuation. The notation ($n,m$) is used here to represent a Fourier mode, where $n/T$ is the frequency and $2\pi m/L_z$ is the spanwise wavenumber of a disturbance wave \cite{23}. Here, $L_z$ is the spanwise domain width. Therefore, mode ($1,0$) represents a 2-D disturbance wave with the fundamental frequency $F=1/T$ and mode ($1,2$) denotes a harmonic wave of period $T$ with spanwise wavelength $\lambda_z=L_z/2$, and so on. The downstream development of the maximum disturbance $u$-velocity Fourier amplitude of a Fourier mode, $A_u$, for the controlled flows with different VGJs scenarios is presented in Figure \ref{fig:Au}. It should be noted that only the modes that reach high amplitudes are highlighted in Figure \ref{fig:Au}. For each case, the streamwise locations of the time- and spanwise-averaged separation and reattachment were identified by vertical dashed lines.

For both cases, a strong peak can be seen in the Fourier spectra which corresponds to mode ($0,1$) representing steady disturbance waves with spanwise wavelength $\lambda_z=L_z$. It appears that this mode is governed by spanwise spacing of VGJs. However, the streamwise growth rate and the onset of the growth associated with this mode is slightly different for two cases. For steady VGJs, mode ($0,1$) decays up to the separation location and then starts to grow further downstream. This mode starts to grow further upstream and experiences stronger growth for harmonic case compare to the steady actuation, hints the fact that different physical mechanism plays a role when harmonic VGJ is employed. Another important observation in Figure \ref{fig:Au}b is that in the steady VGJs case, unsteady disturbances reach appreciable amplitudes only downstream of reattachment location indicating that this flow control strategies is unable to excite fluctuating disturbances. The dominant unsteady Fourier modes for this case are modes ($1,2$) and ($1,3$). This could explain why this type of VGJs forcing is less effective in controlling the separation bubble compared to the harmonic VGJs flow control while injecting more momentum into the flow through the hole. On the other hand, harmonic 2-D and 3-D disturbances with the fundamental frequency $F$, modes ($1,0$), ($1,1$), ($1,2$) and ($1,3$), were strongly excited by harmonic forcing. In particular, mode ($1,1$) is the dominant Fourier mode, which is consistent with the instantaneous flow visualization in Figures \ref{fig:inst_hvgj} where dominant coherent structures with the spanwise wavelength of $\lambda_z=L_z$ were identified.

\section{Conclusions}\label{conc}
High resolution direct numerical simulations of canonical separation bubbles on a flat-plate were carried out for investigating active flow control for a laminar separation bubble using vortex generator jets. The main focus of this paper was to investigate the underlying flow physics of controlled LSBs using steady and harmonic VGJs. From our investigations, we found that both VGJs flow control strategies were successful in reducing the streamwise and wall-normal extent of the separation bubbles when the same blowing amplitude was used for the actuation. However, it was found that harmonic VGJ is indeed more effective in reducing laminar boundary-layer separation than steady forcing whereas larger momentum coefficient was used for the steady VGJ. In particular, the mean bubble length was reduced by 72\% for harmonic VGJs in comparison to 62\% of steady VGJs. Using spectral analysis, the present investigation demonstrated that different physical mechanisms are associated with steady and harmonic VGJs. The effectiveness of steady actuation was related to the generation of steady longitudinal vortices which accelerate the transition process, and as the result, limits the extent of the separated region. In addition to generation of steady modes, harmonic VGJ forcing was able to excite the unsteady 2-D and 3-D disturbances, and the rapid amplification of disturbance waves leads to the generation of strong coherent structures that are very effective in controlling the LSB. While the breakdown to turbulence alone provides a limited amount of momentum exchange, the present results indicated that the generation and development of strong organized coherent structures is primarily responsible for the increased effectiveness of harmonic VGJs compared to steady VGJs.



\end{document}